\documentclass[prd,preprint,superscriptaddress,nofootinbib,longbibliography,aps]{revtex4-1}
\usepackage[utf8]{inputenc}
\usepackage{graphicx}
\usepackage{longtable}
\usepackage{amsmath}
\usepackage{color}
\usepackage{amssymb}
\usepackage{subfigure}
\usepackage{tabularx}
\usepackage{microtype}
 \usepackage[linktoc=all]{hyperref}
 \usepackage{cleveref}
\usepackage{stackengine}
\usepackage[normalem]{ulem}
\usepackage{slashed}
\usepackage{color}
\usepackage{bm}
\usepackage{braket}
\usepackage{slashed}
\usepackage{comment}
\definecolor{myblue}{rgb}{ 0.188, 0.478,0.858}

\begin{document}  
\title{A Test of Gravity with Pulsar Timing Arrays}

\author{Qiuyue Liang}
\email{qyliang@sas.upenn.edu}
\author{Meng-Xiang Lin}
\email{mxlin@sas.upenn.edu}
\author{Mark Trodden} 
\email{trodden@upenn.edu}

\affiliation{Center for Particle Cosmology, Department of Physics and Astronomy, University of Pennsylvania, Philadelphia, Pennsylvania 19104, USA} 

\date{\today}

\begin{abstract}
A successful measurement of the Stochastic Gravitational Wave Background (SGWB) in Pulsar Timing Arrays (PTAs) would open up a new window through which to test the predictions of General Relativity (GR). We consider how these measurements might reveal deviations from GR by studying the overlap reduction function --- the quantity that in GR is approximated by the Hellings-Downs curve --- in some sample modifications of gravity, focusing on the generic prediction of a modified dispersion relation for gravitational waves. We find a distinct signature of such modifications to GR --- a shift in the minimum angle of the angular distribution --- and demonstrate that this shift is quantitatively sensitive to any change in the phase velocity. In a given modification of gravity, this result can be used, in some regions of parameter space, to distinguish the effect of a modified dispersion relation from that due to the presence of extra polarization modes. 
\end{abstract} 

\maketitle

\section{Introduction} 

Efforts to measure the stochastic gravitational wave background (SGWB) using pulsar timing arrays (PTA)~\cite{NANOGrav:2020bcs,Goncharov:2021oub,Chen:2021rqp,Antoniadis:2022pcn} hold out the promise of adding a novel dimension to the newly-arrived epoch of gravitational wave astrophysics and cosmology~\cite{Bailes:2021tot,EventHorizonTelescope:2019dse,Johannsen:2015hib}. Pulsars are rapidly rotating sources, whose beamed emissions sweep the earth with extremely precise periodicities. Because the rotational periods of pulsars --- especially millisecond pulsars --- are so intrinsically stable, high precision measurements of the arrival times of the pulse signals can serve as a sensitive probe of fundamental physics and, in particular, of gravity. More specifically, any gravitational waves passing through the spacetime between the pulsar and earth will, by definition, perturb the metric, and hence affect the arrival times of pulses. Therefore, monitoring the pulsar signal arrival times using PTA measurements can in principle be used to detect the SGWB. In General Relativity, the angular dependence of these signals is captured by an angular correlation function known as the Hellings-Downs curve \cite{Hellings:1983fr}. Thus, a successful detection of gravitational wave signals vias PTAs opens up the possibility that deviations from this curve might be observed, pointing to new physics beyond GR. 

There are a number of different ways in which PTA measurements may constrain modifications to GR. In particular, such deviations may occur because in more general theories there are up to six polarization modes that can influence PTA results: two transverse-traceless tensor modes, two vector modes, one scalar-longitudinal mode, and one scalar-transverse mode. Indeed, a number of authors have already explored deviations from the predictions of GR in PTA measurements; for example, see \cite{Lee:2010cg,Gair:2015hra,Qin:2020hfy,Liang:2021bct,Bernardo:2022rif}.

In this paper, we focus on tensor modes, as we expect these to be the most generic signal, since scalar modes, for example, can be screened in some theories~\cite{Joyce:2014kja,deRham:2012fw,Chu:2012kz,deRham:2012fg,Dar:2018dra}. We study the angular correlations of PTA measurement of SGWB in theories with general modified dispersion relations, by employing the spherical harmonics decomposition method that is widely used in the analysis of the cosmic microwave background (CMB), and which can be applied effectively to PTA measurements~\cite{Gair:2014rwa}. We demonstrate that modifications to GR change the relative contributions of different multipoles, and hence change the shape of the overlap reduction function. In particular, the angle of the minimum of the overlap reduction function shifts in a predictable way as a response to the change in the phase velocity of the GWs. These features can be used to distinguish among different types of gravitational theories using future PTA datasets.

The paper is organized as follows. In Sec.~\ref{sec:decom}, we briefly review the formalism of the spherical harmonics decomposition. In Sec.~\ref{sec:PTA}, we present general calculations for the angular correlation function of the SGWB in PTA observations, and in Sec.~\ref{sec:example} we investigate two concrete examples --- massive gravity and the case of gravity with a  subluminal phase velocity. We conclude in Sec.~\ref{sec:conclude}. Throughout we use the metric signature $(-,+,+,+)$.

\section{General Formalism of the Spherical Harmonic Decomposition} \label{sec:decom}
It is particularly convenient for the calculations in this paper to work in the formalism~\cite{Gair:2014rwa}, in which the gravitational wave background is decomposed into spherical harmonics in the same way that is traditionally applied to the polarizations of the CMB. In this section we briefly collect the relevant definitions and review how this formalism is applied in our case.

A gravitational wave is a tensor perturbation $h_{\mu\nu}(t,{\bf x})$ of the metric, which can be written in terms of its spatial components $h_{ij}(t,{\bf x})$, decomposed into $+$ and $\times $ polarization modes. If we write the direction of propagation as $\hat \Omega$, and Fourier transform with respect to time, so that we are in frequency space, this decomposition can be written as 
\begin{equation}
    h_{ij}(f,\hat\Omega) = h^+(f,\hat\Omega) {\tilde e}^+_{ij}(\hat \Omega)+ h^\times(f,\hat\Omega) {\tilde e}^\times _{ij} (\hat \Omega) \ ,
\end{equation}
where
 \begin{eqnarray}
\label{eq,polarization}
 &&{\tilde e}_{i j}^{+}(\hat{\Omega})=\hat{m}_i \hat{m}_j-\hat{n}_i \hat{n}_j  
 \ ,{\tilde e}_{i j}^{\times}(\hat{\Omega})=\hat{m}_i \hat{n}_j+\hat{n}_i \hat{m}_j \nonumber\\ 
 && \hat\Omega= (\sin \theta \cos \phi, \sin \theta \sin \phi, \cos \theta) \nonumber\\
 && \hat{m}=(\sin \phi,-\cos \phi, 0)\ , \hat{n}=(\cos \theta \cos \phi, \cos \theta \sin \phi,-\sin \theta) \ .
\end{eqnarray} 
Changing coordinates from Cartesian ones to spherically symmetric ones via $\hat\Omega = \hat r$, $\hat m = \hat \phi$, $\hat n = \hat \theta $, the gravitational wave can be projected onto the 2-sphere with a fixed radius. This allows us to express the gravitational wave in a two-dimensional space as $h_{ab}(f,\hat\Omega)$ by defining projected polarization tensors 
\begin{eqnarray}
\label{eq,polarizationbasis}
   e_{ab}^{+} = \hat \phi_a \hat\phi_b - \hat\theta_a\hat\theta_b, \quad  e_{ab}^{\times} = \hat \phi_a \hat\theta_b + \hat \theta_a\hat\phi_b  \ ,
\end{eqnarray}
 where the indices $a,b, \dots$ run over $(\theta,\phi)$.
 
As is well-known in CMB physics, there exists another useful basis into which we may decompose rank-2 tensors on the 2-sphere. We write the gradient and curl respectively of the spherical harmonics, $Y_{(lm)}$, as
\begin{eqnarray}
\label{eq,harmonicbasis}
    Y_{(\ell m) a b}^E=N_l\left(Y_{(\ell m) ; a b}-\frac{1}{2} g_{a b} Y_{(\ell m) ; c}{ }^c\right)\ , \ \ \  Y_{(\ell m) a b}^B=\frac{N_\ell}{2}\left(Y_{(\ell m) ; a c} \epsilon_b^c+Y_{(\ell m) ; b c} \epsilon_a^c\right)\ ,
\end{eqnarray}
where $\epsilon_{ab}$ is the Levi-Civita symbol, and 
\begin{equation}
    N_\ell = \sqrt{2\frac{(\ell-2)!}{(\ell+2)!}}.
\end{equation} 
In this basis, a gravitational wave can then be expressed as
\begin{eqnarray}\label{eq:h_Ylm}
    h_{ab} (f, \hat\Omega) = \sum_{\ell=2}^{\infty} \sum_{m=-\ell}^l\left[a_{(\ell m)}^E(f) Y_{(\ell m) a b}^E(\hat\Omega)+a_{(\ell m)}^B(f) Y_{(\ell m) a b}^B(\hat\Omega)\right] \ .
\end{eqnarray}

These two bases are related by 
\begin{eqnarray}\label{eq,decom}
    &&  Y_{(\ell m) a b}^{E} (\hat\Omega)=\frac{N_\ell}{2}\left[W_{(\ell m)}(\hat\Omega) e_{a b}^{+}(\hat\Omega) + X_{(\ell m)}(\hat\Omega) e_{a b}^{\times}(\hat\Omega)\right] , \\
    &&  Y_{(\ell m) a b}^{B} (\hat\Omega)=\frac{N_\ell}{2}\left[W_{(\ell m)}(\hat\Omega) e_{a b}^{\times}(\hat\Omega) - X_{(\ell m)}(\hat\Omega) e_{a b}^{+}(\hat\Omega)\right] , 
\end{eqnarray}
with the coefficients being related by
\begin{eqnarray}
\label{eq,relation1}
    && h^{+}(f, \hat\Omega)=\sum_{(\ell m)} \frac{N_\ell}{2}\left[a_{(\ell m)}^E(f) W_{(\ell m)}(\hat\Omega)-a_{(\ell m)}^B(f) X_{(\ell m)}(\hat\Omega)\right], \\
    && h^{\times}(f, \hat\Omega)=\sum_{(\ell m)} \frac{N_\ell}{2}\left[a_{(\ell m)}^E(f) X_{(\ell m)}(\hat\Omega)+a_{(\ell m)}^B(f) W_{(\ell m)}(\hat\Omega)\right].
\end{eqnarray}
Conversely
\begin{eqnarray}
\label{eq,relation2}
    && a_{(\ell m)}^E(f) = N_\ell \int d^2\hat\Omega \left[h^{+}(f,\hat\Omega)W^*_{(\ell m)}(\hat\Omega) +h^{\times}(f,\hat\Omega)X^*_{(\ell m)}(\hat\Omega) \right], \\
    && a_{(\ell m)}^B(f) = N_\ell \int d^2\hat\Omega \left[h^{\times}(f,\hat\Omega)W^*_{(\ell m)}(\hat\Omega) -h^{+}(f,\hat\Omega)X^*_{(\ell m)}(\hat\Omega) \right] ,
\end{eqnarray}
and 
\begin{eqnarray}
\label{eq,associated}
    && W_{(\ell m)}(\hat\Omega) = \left( \frac{\partial^2}{\partial\theta^2}-\cot\theta+\frac{m^2}{\sin^2\theta}\right) Y_{(\ell m)}(\hat\Omega) = \left(2\frac{\partial^2}{\partial\theta^2}+l(l+1)\right) Y_{(\ell m)}(\hat\Omega),\\
    && X_{(\ell m)}(\hat\Omega) = \frac{2im}{\sin\theta} \left(\frac{\partial}{\partial\theta}-\cot\theta\right)Y_{(\ell m)}(\hat\Omega).
\end{eqnarray}
Here, $W_{(\ell m)}$ and $X_{(\ell m)}$ are related to the associated Legendre polynomials~\cite{Gair:2014rwa}.

Assuming that there are no correlations between different frequencies in the stochastic background, the power spectrum can then either be represented by 
\begin{eqnarray}\label{eq:H(f)}
    \left\langle a_{(\ell m)}^E(f) a_{\left(\ell^{\prime} m^{\prime}\right)}^{E *}\left(f^{\prime}\right)\right\rangle =\left\langle a_{(\ell m)}^B(f) a_{\left(\ell^{\prime} m^{\prime}\right)}^{B *}\left(f^{\prime}\right)\right\rangle =  H(f) \delta_{\ell \ell^\prime}\delta_{m m^\prime}   \delta\left(f-f^{\prime}\right)  \ ,
\end{eqnarray}
or by
\begin{eqnarray}\label{eq:H(f)2}
    \left\langle h^+(f,\hat \Omega ) h^{+ *}(f^{\prime},\hat \Omega^\prime )\right\rangle =\left\langle h^\times(f,\hat \Omega ) h^{\times  *}(f^{\prime},\hat \Omega^\prime )\right\rangle  = \frac{1}{2} H(f) \delta^2 (\hat \Omega,\hat\Omega^\prime )  \delta\left(f-f^{\prime}\right)  \ .
\end{eqnarray} 

In the next section, we will define the observables relevant for PTA measurements, and will then use the decomposition above to extract these observables from measurements of the timing residuals.

\section{Pulsar Timing Arrays and Modified Gravity} \label{sec:PTA}

We now consider the PTA detector response in the framework of modified gravity. We shall see below that the result is determined by the dispersion relation satisfied by the gravitational waves. Therefore, in the following,  without loss of generality, we will encode the effects of deviations from GR in a general dispersion relation $\omega(k)$. As is usual, we assume that the stochastic gravitational wave signals can be approximated as plane waves. This approximation is valid when the signal from a single source of the stochastic background can be adiabatically approximated as monochromatic as a function of time (as during the inspiral phase, but not the merger phase, of binary systems). One example of this is the inspiral signal from binary coalescence, which is expected to be the major contribution to PTA SGWB detection, although the assumption may hold more broadly for other contributors from physics in the early universe. Under this assumption, the results only depend on the phase velocity of the gravitational wave rather than the group velocity, as we shall see below.

For an individual pulsar, the important quantity is the residual of the pulse arrival time,
\begin{equation}
\label{eq,R}
R(t) \equiv \int_0^t dt^\prime\, \left(\frac{\nu_0-\nu(t^\prime)}{\nu_0}\right) \ ,
\end{equation}
where $\nu_0$ is the frequency of the pulse in a flat spacetime, $\nu$ is the perturbed frequency due to the underlying metric perturbation, i.e. gravitational waves, and where we have suppressed any dependence on the direction of the pulsar until we explicitly need to retain it when dealing with multiple sources. PTA observations study the correlations between measurements of this quantity for different pulsars in order to increase the signal-to-noise ratio. 

To calculate this observable, we start from null geodesics in the perturbed spacetime, which are given by parametrized null vectors $\sigma^\mu (\lambda)$, constructed from null geodesics in Minkowski spacetime, $s^\mu \equiv \nu (1,-{\hat p} ) = dx^\mu/d\lambda $ via 
\begin{equation}
 \sigma^\mu (\lambda) = s^\mu(\lambda)- \frac{1}{2} \eta^{\mu\nu} h_{\nu \gamma}(x(\lambda) )  s^{\gamma} (\lambda)\ .
 \label{eq:pertnullgeo}
\end{equation}
These are the paths taken by light from the pulsar to Earth.

Using the geodesic equation, we may relate the change in pulse frequency to the metric perturbation $h_{\mu\nu}$, via
\begin{equation}
    \label{eq,geodesic}
\frac{d \sigma^{\mu}}{d \lambda} =  -\Gamma_{\alpha \beta}^{\mu} \sigma^{\alpha} \sigma^{ \beta}=  -\frac{1}{2} \eta^{\mu\nu} \left(h_{\nu\alpha,\beta}+h_{\nu\beta,\alpha} -h_{\alpha\beta,\nu} \right) s^\alpha s^\beta + \mathcal{O}(h_{\mu\nu}^2) \ .
\end{equation}
Since we only study tensor modes, as discussed in the previous section, we focus on the spatial components of the metric perturbation. This yields,
\begin{equation}
\label{eq,s0}
 \frac{ds^0}{d\lambda} = \frac{1}{2} \dot h_{ij} s^i s^j +\mathcal{O}(h^2) \ ,
\end{equation}
as in the GR case. However, since we are allowing for the possibility that the dispersion relation differs from that in GR, the mode function of the plane wave is given by
\begin{equation}
\label{eq,fourierhA}
 h_{ij } \left(  t-\frac{1}{v_{ph}}\hat{ {\Omega}}  \cdot   \vec{x} \right) = \int_{-\infty}^\infty d f  ~ e^{i 2\pi f \left(  t- \frac{1}{v_{ph}} \hat{ \Omega }  \cdot   \vec{x}\right) }  h_{ij }\left(f, \frac{1}{v_{ph}} \hat{ \Omega }\right)  \ ,
\end{equation}
where $v_{ph}\equiv \omega (k)/k$ is the phase velocity which, under the assumption of plane waves, encodes the deviation from GR. When $v_{ph}=1$ this expression reduces to the GR case.
In the case of massive gravity, for example, we have $\omega^2 = k^2 + m^2$, and thus $v_{ph}>1$; while in the case of a dispersion relation $\omega = c_s k$ with sound speed $c_s<1$, we instead have $v_{ph}<1$.

We can express $dh_{ij}/d\lambda$ as 
\begin{equation}
    \frac{d h_{ij} (  t-\frac{1}{v_{ph}} \hat{\Omega} \cdot \vec x)}{d\lambda} =\frac{\partial h_{ij}}{\partial x^0 } \frac{d x^0}{d\lambda} + \frac{\partial h_{ij}}{\partial\hat{\Omega}  \cdot \vec x } \frac{d \hat{\Omega} \cdot \vec x }{d\lambda} = \frac{\partial h_{ij}}{\partial x^0} \left(  \frac{dx^0 }{d\lambda} - \frac{1}{v_{ph}}  \hat{\Omega} \cdot \frac{d\vec x }{d\lambda} \right) = \dot h_{ij} \nu \left(1 + \frac{1}{v_{ph}}  \hat{\Omega} \cdot \hat{p} \right) \ ,
\end{equation}
where $d\vec x/d\lambda = -\nu\hat{ p}$ is the spatial momentum of the pulsar signal. Substituting this expression into Eq.\eqref{eq,s0}, and using that $s^0=\nu$ we then obtain,
\begin{eqnarray}
\label{eq,s0final}
  \frac{d\nu}{d\lambda} =  \frac{ \nu }{2 \left(1 +\frac{1}{v_{ph}} \hat{\Omega}   \cdot \hat{p} \right) } \frac{d h_{ij }}{d\lambda} \hat p^i \hat p^j  \ .
\end{eqnarray}
Integrating this quantity along pulsar-Earth path yields,
\begin{equation}
    \log\left(\frac{\nu(t)}{\nu_0}\right) =    \frac{ \hat p^i \hat p^j  }{2  \left(1+ \frac{1}{v_{ph}} \hat{\Omega}   \cdot \hat{p} \right)  }  \Delta h_{ij } \ ,
\end{equation}
where
\begin{equation}\label{eq,differentmetric}
    \Delta h_{ij } \equiv h_{ij }\left(t_{\mathrm{p}}, \frac{1}{v_{ph}} \hat{\Omega}\right)-h_{ij }\left(t_{\mathrm{e}}, \frac{1}{v_{ph}} \hat{\Omega}\right)
\end{equation}
is the difference between the metric perturbation at the pulsar, and that received at Earth after traveling along the direction $\hat{\Omega} $. Exponentiating both sides and expanding to $\mathcal{O}(h)$, we can then define the {\it redshift} as the fractional change in frequency via,
\begin{eqnarray}\label{eq, nu2}
    z \equiv \frac{\nu_0 - \nu(t)}{\nu_0}  &=&  - \frac{1}{2}\frac{ \hat p^i \hat p^j  }{\left(1  +\frac{1}{v_{ph}} \hat{\Omega}   \cdot \hat{p} \right)  }  \Delta h_{ij }  = -\frac{1}{2}\frac{ \hat p^a \hat p^b  }{\left(1  +\frac{1}{v_{ph}} \hat{\Omega}   \cdot \hat{p} \right)  }  \Delta h_{ab }    \ ,
\end{eqnarray}
where in the final step. as discussed in the previous section, we have projected the gravitational wave onto the 2-sphere with a fixed radius and parametrized using spherical coordinates.

It is convenient to choose a coordinate system in which the distance between a pulsar and Earth is denoted by $L$, and we write~\cite{Anholm:2008wy} (note, the photon is massless and travels at the speed of light), 
\begin{equation}
    \vec x_e = 0, \quad \vec x_p = L \hat{p} ,
    \quad  t_p = t_e -L \equiv t-L \ .
\end{equation}
We use the Fourier transformation defined in Eq.\eqref{eq,fourierhA} to express the difference between metric perturbations Eq.\eqref{eq,differentmetric} (projected onto the 2-sphere) as
\begin{eqnarray}
    \begin{aligned}
    \Delta h_{ab }=\int_{-\infty}^{\infty} d f & e^{i 2 \pi f t}\left(e^{-i 2 \pi f L\left(1 +\frac{1}{v_{ph}} \hat{\Omega}   \cdot \hat{p} \right)}-1\right) \times h_{ab }\left(f, \frac{1}{v_{ph}} \hat{\Omega} \right) \ .
    \end{aligned}
\end{eqnarray}
 
Since the effect of the stochastic gravitational wave background on the measured redshift of a given pulsar consists of contributions from gravitational waves arriving from all directions, a relevant quantity to calculate is the total redshift
\begin{equation}\label{eq,redshift}
    \tilde z (f,\hat{p}) \equiv \int_{S^2 } d^2 \hat{\Omega} ~ z (f,\hat{p},\hat{\Omega}  ) = \int_{S^2 } d^2 \hat{\Omega} ~   \left(1- e^{-i 2 \pi f L\left(1+\frac{1}{v_{ph}} \hat{\Omega} \cdot \hat{p}\right)}\right) \frac{\hat{p}^a \hat{p}^b}{2(1+\frac{1}{v_{ph}} \hat{\Omega} \cdot \hat{p})} h_{ab}\left(f, \frac{1}{v_{ph}} \hat{\Omega}\right) \ .
\end{equation}
Although we explicitly retain the $\hat p$ dependence in order to keep track of the angular dependence of the pulsar term, one should keep in mind that we will eventually integrate over the direction of the gravitational wave, $\hat \Omega$, since we are interested in the stochastic gravitational wave background. The timing residual Eq.\eqref{eq,R} is then given by
\begin{eqnarray}\label{eq,residue}
    R(t, \hat p) = \int_0^t z(t^\prime) dt^\prime = \int_{-\infty}^\infty \frac{df e^{2\pi i ft}}{2\pi i f} \int_{S^2} d^2\hat\Omega \frac{\hat p^a\hat p^b h_{ab}}{2(1+ \frac{1}{v_{ph}} \hat\Omega\cdot\hat p)} \left(1- e^{-2\pi i fL \left(1+ \frac{1}{v_{ph}} \hat\Omega\cdot \hat p\right) }\right) \ ,
\end{eqnarray} 
from which we can read off the Fourier transform $\tilde R(f,\hat p)$ as 
\begin{eqnarray}
    \tilde R(f, \hat p) &=&  \frac{1}{2\pi i f} \int_{S^2}d^2\hat\Omega \frac{\hat p^a\hat p^b }{2(1+ \frac{1}{v_{ph}} \hat\Omega\cdot\hat p)} \left(1- e^{-2\pi i fL \left(1+ \frac{1}{v_{ph}} \hat\Omega\cdot \hat p\right) }\right) h_{ab} \left(f, \frac{1}{v_{ph}} \hat\Omega \right)\nonumber\\
    &\equiv &    \int_{S^2}d^2\hat\Omega R^{ab} (f, \hat p,\hat \Omega)h_{ab}\left(f, \frac{1}{v_{ph}} \hat\Omega \right)  \ ,
\end{eqnarray}
where $R^{ab}(f, \hat p,\hat \Omega) $ is the {\it detector response function}, a rank-2 tensor living on the 2-sphere. The detector response function can be decomposed either into the basis of harmonics, $Y_{(\ell m)ab}^{E,B}$ as in Eq.\eqref{eq,harmonicbasis}, 
\begin{eqnarray}
    R_{ab}(f, \hat p,\hat\Omega) = \sum_{\ell=2}^\infty \sum_{m=-\ell}^\ell \left(R^{ E}_{(\ell m)}(f, \hat p ) Y^E_{(\ell m) ab }(\hat\Omega) +R^{  B}_{(\ell m)}(f, \hat p ) Y^B_{(\ell m) ab } (\hat\Omega)   \right) \ ,
\end{eqnarray}
or can be decomposed into the basis of polarization tensors, $e_{ab}^{+,\times}$ as in Eq.\eqref{eq,polarizationbasis}
\begin{eqnarray}
    R_{ab}(f, \hat p,\hat\Omega) =  R^{+}(f, \hat p,\hat\Omega) e_{ab}^+ (\hat\Omega) +  R^{\times}(f, \hat p,\hat\Omega) e_{ab}^\times (\hat\Omega)  \ .
\end{eqnarray}
The coefficients of these two equivalent decompositions are related to each other in a similar way to the relationship between equivalent decompositions of the graviton, as in Eq.\eqref{eq,relation1} and \eqref{eq,relation2}.

Making use of the orthogonality properties of these bases, we can then represent $\tilde R(f,\hat p)$ as 
\begin{eqnarray}
\label{eq,tildeRf}
    \tilde R(f,\hat p) = \sum_{\ell=2}^\infty \sum_{m= -\ell}^\ell \left(R^E_{(\ell m)}(f,\hat p) a_{(\ell m)}^E(f) + R^B_{(\ell m)}(f,\hat p) a_{(\ell m)}^B(f) \right) \ , 
\end{eqnarray}
or
\begin{eqnarray}
\label{eq,tildeRf2 }
    \tilde R(f,\hat p) =\int_{S^2} d^2\hat\Omega \left(R^+ (f,\hat p,\hat \Omega) h ^+(f,\hat \Omega ) +R^\times (f,\hat p,\hat \Omega) h ^\times (f,\hat \Omega) \right) \ .
\end{eqnarray}
In the balance of this paper, we will refer to these coefficients $R_{(\ell m)}^P$ (with $P=E,B$), and $R^A$ (with $A=+,\times$) as the {\it response functions}. For example, we will refer to $R_{(\ell m)}^E$ as the E-mode response function, and $R^+$ as the $+$ mode response function. 

For the PTA system, it is possible to obtain the specific forms of the $R_{(\ell m)}^P(f,\hat{p})$:
\begin{equation}
\label{eq,RlmP}
    R_{(\ell m)}^P(f,\hat{p})=\frac{1}{2 \pi i f} \int_{S^2} \mathrm{d}^2 \hat\Omega  \frac{\hat{p}^{a} \hat{p}^b   }{2\left(1+\frac{1}{v_{ph}} \hat{\Omega} \cdot \hat{p}\right)} ~ Y_{(\ell m) a b}^P(\hat{\Omega})  \left[1-e^{-i 2 \pi f L\left(1+\frac{1}{v_{ph}}\hat{\Omega} \cdot \hat{p}\right) }\right] \ .
\end{equation}
Our goal is to compute the correlation function of the timing residuals. If we assume that the stochastic gravitational wave background is isotropic, the dependence of this quantity on the angle between pairs of pulsars, $\xi = \cos^{-1}(\hat p_1 \cdot \hat p_2)$, can be factorized from the dependence on the power spectrum of SGWB. We can then express the correlation function for a pair of pulsars as
\begin{eqnarray}
\label{eq,twopt}
    \braket{R(t, \hat p_1 ) R(t^\prime,\hat p_2)  } = \int_{-\infty}^\infty df e^{2\pi i f(t- t^\prime )} H(f) \Gamma(f,\xi)\ ,
\end{eqnarray}
where, $H(f)$ (defined in Eq.\eqref{eq:H(f)} and \eqref{eq:H(f)2}), encodes information about the power spectrum of the SGWB, and $\Gamma(f,\xi)$, which is known as the {\it overlap reduction function}, contains information about the angular distribution. If we choose the harmonic basis, and use Eq.\eqref{eq:H(f)} and \eqref{eq,tildeRf}, we see the overlap reduction function can be calculated as 
\begin{eqnarray}
\label{eq,gammasum}
    \Gamma(f,\xi) =\mathcal{C} \sum_{\ell=2}^{\infty} \Gamma_{12, \ell}  (f,\xi) = \mathcal{C}  \sum_{\ell=2}^{\infty} \sum_{m= -\ell}^{\ell} \sum_{P= E,B } R_{(\ell m)}^P (f,\hat{p}_1)  R_{(\ell m)}^{P*} (f,\hat{p}_2)  \ ,
\end{eqnarray}
where the dependence on $\xi$ arises from the $\hat p$ dependence of $R_{(\ell m)}^P(f,\hat p) $, and $\mathcal{C}$ is the normalization factor.
Note that, equivalently, we are free to choose the polarization basis, and to use Eq.\eqref{eq:H(f)2} and \eqref{eq,tildeRf2 } to express the overlap reduction function as 
\begin{eqnarray}
\label{eq,gammaint}
    && \Gamma(f,\xi) = \beta_T\int_{S^2}d^2 \Omega \sum_{A= +,\times } R^A (f,\hat{p}_1,\hat\Omega )  R^{A*}  (f,\hat{p}_2,\hat\Omega )   \nonumber\\
    &=& \beta_T\int_{S^2}d^2 \Omega \left(1-e^{-2 \pi i f L\left(1+\frac{1}{v_{p h}} \hat{\Omega} \cdot \hat{p}_1\right)}\right)\left(1-e^{2 \pi i f L\left(1+\frac{1}{v_{p h}} \hat{\Omega} \cdot \hat{p}_2\right)}\right) \nonumber\\
    &&\times \sum_{A=+, \times}\left(\frac{\hat{p}_1^i \hat{p}_1^j e_{i j}^A}{2\left(1+\frac{1}{v_{p h}} \hat{\Omega} \cdot \hat{p}_1\right)} \frac{\hat{p}_2^i \hat{p}_2^j e_{i j}^A}{2\left(1+\frac{1}{v_{p h}} \hat{\Omega} \cdot \hat{p}_2\right)}\right) \ .
\end{eqnarray} 
This is the more familiar expression for the overlap reduction function, often seen in the literature, where $\beta_T$ is the normalization factor.  As we have seen, the different choices for the decomposition are equivalent, and we are free to use whichever is the most convenient for the task at hand.

Now, for our purposes it is convenient to use the harmonic basis to compute the exact form of the overlap reduction function using Eq.\eqref{eq,gammasum} and \eqref{eq,RlmP}. We first note that we are always free to choose a frame in which the pulse $\hat p_I $ is aligned with the $z-$axis~\footnote{For an arbitrary pulsar direction with angular coordinates, $\hat p_I = \left(\zeta_I, \chi_I\right)$, we can use the rotation matrix $ \mathbf{R}(\chi_I,\zeta_I ,0 ) \hat p_I = \bar{\hat p}_I = (0,0,1)$ to carry out this transformation.}, so that $\hat p_I = \bar{\hat p}_I = (0,0,1)$. Under such a rotation, the direction of the gravitational wave is given by $\bar{\hat\Omega}$, with $\bar{\hat\Omega} \cdot \bar{\hat p} = \cos\bar\theta$. 
The gradient and curl of the spherical harmonics then transform under the rotation as:
\begin{eqnarray}
    Y_{(\ell m) a b}^P(\theta, \phi)=\sum_{m^{\prime}=-\ell}^\ell\left[D^\ell{ }_{m m^{\prime}}(\chi, \zeta, 0)\right]^* Y_{\left(\ell m^{\prime}\right) \bar{a} \bar{b}}^P\left(\bar{\theta}_I, \bar{\phi}_I\right) \mathbf{R}\left(\chi_I, \zeta_I, 0\right)^{\bar{a}}{ }_a \mathbf{R}\left(\chi_I, \zeta_I, 0\right)^{\bar{b}}{ }_b \ ,
\end{eqnarray}
where $D^\ell{ }_{m m^{\prime}} $ is the Wigner-D matrix associated with the rotation matrix. Therefore, the E-mode response function is given by
\begin{eqnarray}
R_{(\ell m)}^E(f,\hat{p}) &=& \frac{1}{2\pi i f} \int_{-1}^{1} d\cos\theta \int_0^{2\pi} d\phi \frac{1}{2}\frac{\hat p^a\hat p^b}{1+ \frac{1}{v_{ph}} \hat\Omega \cdot \hat{p} }  Y_{(\ell m)ab}^E ( \theta,\phi ) \left[1-e^{-i 2 \pi f L(1+\frac{1}{v_{ph}} \hat{\Omega } \cdot \hat{p}) }\right] \nonumber\\
&=& \sum_{m^{\prime}=-\ell}^\ell\left[D^\ell{ }_{m m^{\prime}}(\chi, \zeta, 0)\right]^*    \frac{1}{2\pi i f} \nonumber\\
    &&\times \int_{-1}^{1} d\cos\bar \theta \int_0^{2\pi} d\bar \phi \frac{1}{2}\frac{\bar p^a \bar p^b}{1+\frac{1}{v_{ph}} \cos\bar\theta }  Y_{(\ell m^{\prime})ab}^E (\bar\Omega ) \left[1-e^{-i 2 \pi f L(1+\frac{1}{v_{ph}} \cos\bar \theta ) }\right]  \nonumber\\
&=&   \sum_{m^{\prime}=-\ell}^\ell\left[D^\ell{ }_{m m^{\prime}}(\chi, \zeta, 0)\right]^*    \frac{1}{2\pi i f} \nonumber\\
    &&\times \int_{-1}^{1} d\cos\bar \theta \int_0^{2\pi} d\bar \phi    \left[1-e^{-i 2 \pi f L(1+\cos\bar \theta ) }\right]  \frac{N_l}{2}\left( W_{\ell m^{\prime}} F^+ + X_{\ell m^{\prime}} F^\times  \right) \ ,
\end{eqnarray}
where in the last line we have used Eq.\eqref{eq,decom}, and have defined 
\begin{eqnarray}
    F^{+,\times}(\bar\Omega) = \frac{1}{2}\frac{\bar p^a \bar p^b}{1+ \frac{1}{v_{ph}} \cos\bar\theta }  e^{+,\times}_{ab}(\bar\Omega) \ .
\end{eqnarray}
In the frame $\bar p = (0,0,1)$, we have $F^+ =  (1-\cos^2\bar\theta)/2(1+\frac{1}{v_{ph}} \cos\bar\theta  )$, and $F^\times = 0$. Moreover, since $F^+$ does not depend on $\bar\phi$, we conclude that the only contribution in the summation is from the term with $m^{\prime} = 0$. Therefore, $R_{(\ell m)}^{E}$ is
\begin{eqnarray}
\label{eq:RlmE}
    R_{(\ell m)}^E(f,\hat{p})=\frac{\left[D_{m 0 }^\ell(\chi, \zeta, 0)\right]^*   }{ 2 \pi i  f}  \frac{N_\ell}{2}\int d^2\hat\Omega \left[1-e^{-i 2 \pi f L\left(1+ \frac{1}{v_{ph}} x \right)}\right] \frac{1-x^2}{1+\frac{1}{v_{ph}} x } W_{(\ell 0)}(x)\ ,
\end{eqnarray}
where here, and in what follows, we have written $x = \cos\bar\theta$ for simplicity. 
Note that in this frame $F^{\times}(\bar\Omega)=0$ and $X_{l0}(\bar\Omega) =0$, so that it can be readily verified that $R_{(\ell m)}^B$ always vanishes.

The form of the associated Legendre polynomials can be found in \cite{Gair:2014rwa} and, as we review in the next section, in the case of GR the integration has a nice analytic form in the approximation that we may drop the exponential factor. As we will also see, away from the GR limit it is more challenging to work with this approximation. For now we merely point out that in Eq.\eqref{eq:RlmE} the integrand only depends on the mode $\ell$, and so we can isolate the important part of the integral by defining 
\begin{eqnarray}
\label{eq,coeff}
    c_\ell (f) &\equiv & \sqrt{\frac{4}{(2l+1)\pi}}  \int d^2\hat\Omega \left[1-e^{-i 2 \pi f L\left(1+\frac{1}{v_{ph}}  x\right)}\right] \frac{1-x^2}{1+x/v_{ph}} W_{(\ell 0)}(x)  \nonumber \\
        &=& \int_{-1}^1 dx \frac{(1-x^2)^2} { 1+  x/v_{ph} } \left[1-e^{-i 2 \pi f L\left(1+\frac{1}{v_{ph}}  x\right)}\right] \frac{d^2}{dx^2 } P_\ell(x )\ ,
\end{eqnarray}
so that Eq.\eqref{eq:RlmE} reads
\begin{eqnarray}
    R_{(\ell m)}^E(f,\hat{p}) = \frac{\left[D_{m 0}^\ell(\chi, \zeta, 0)\right]^* }{2\pi i f}\frac{N_\ell   }{2 } \sqrt{\frac{(2\ell+1)\pi }{4} } c_\ell(f)=  \frac{Y_{\ell m}(\chi,\zeta)}{2\pi i f}\frac{N_\ell \pi  }{2 }  c_\ell (f)\ .
\end{eqnarray} 
In this last expression we have used the relationship
\begin{eqnarray}
 \left[D_{m 0}^\ell(\chi, \zeta, 0)\right]^* = \sqrt{\frac{4\pi}{2\ell+1}   }Y_{\ell m}(\chi,\zeta)
\end{eqnarray}
between Wigner $D$ matrices and spherical harmonics. The overlap reduction function of mode $\ell$ can therefore be written as 
\begin{eqnarray}
    \Gamma_{12,\ell}(f,\xi) &=& \frac{1}{(2\pi f)^2 }  \sum_{m=-\ell}^\ell    Y_{\ell m}(\chi_1,\zeta_1)Y_{\ell m}^* (\chi_2,\zeta_2)  \left(\frac{N_\ell \pi   }{2 } \right)^2 |c_\ell (f)|^2  \nonumber\\
    &=&  \frac{1}{(2\pi f)^2 }  \frac{2\ell+1}{4\pi} \left(\frac{N_\ell \pi   }{2 } \right)^2 |c_\ell (f)|^2  ~ P_\ell(\cos\xi)\ ,
\end{eqnarray}
where we have used the addition formula of spherical harmonics. After absorbing a factor of  $1 / (64\pi f ^2)$ into the definition of the power spectrum $H(f)$, we then obtain a Legendre polynomial decomposition for the overlap reduction function 
\begin{equation}
\label{eq, overlapnew}
    \Gamma(f,\xi) = \mathcal{C}\sum_{\ell=2}^{\infty} a_\ell P_\ell(\cos\xi)\equiv \mathcal{C}\sum_{\ell=2}^{\infty} (2\ell+1)\frac{2 (\ell-2)!}{(\ell+2)!} |c_\ell(f)|^2 P_\ell(\cos\xi)\ ,
\end{equation}
where $\mathcal{C}$ is an overall factor corresponding to the normalization.

We have gone to great pains to lay out a general approach to the computation of the overlap reduction function because we wish to understand it in situations beyond GR, in which some of the more familiar approaches may no longer hold. However, it is worth briefly reviewing what happens in the case of GR, to understand that point of comparison. As discussed in \cite{Anholm:2008wy}, in GR ($v_{ph}=1$), the exponential factor in the overlap reduction function leads to a damping oscillation. For $f L \gtrsim 10$, it is a good approximation to drop this exponential, and the remaining integral is then straightforward, yielding the well-known {\it Hellings-Downs} curve
\begin{eqnarray}
\label{eq,overlapHD}
    \Gamma_{\text{HD}}(\xi )&=& \frac{\beta_T}{4} \frac{2 \pi}{3}\left(3+\cos \xi+6(1-\cos \xi) \log \frac{1-\cos \xi}{2}\right) \nonumber\\
    &=& \mathcal{C}\sum_{\ell=2}^\infty    \frac{2 (2\ell +1) }{(\ell+2)(\ell+1) \ell (\ell-1)}  P_\ell(\cos \xi ) \ ,
\end{eqnarray}
where, in this approximation, there is no frequency-dependence. As we will review in the next section, this method can be straightforwardly generalized to the massive gravity case, where the exponential term can be ignored and an analytical expression can be obtained, as discussed in \cite{Liang:2021bct}.   

A particular advantage of the harmonic method is that it makes the composition of the overlap reduction function considerably more transparent. For example, as one can easily recognize in Eq.\eqref{eq, overlapnew}, the contribution to the overlap reduction function from tensor modes does not contain $\ell=0$ and $\ell=1$ modes, corresponding to vanishing monopole and dipole contributions in the overlap reduction function, even with a modified dispersion relation. 
These results are far harder to extract from a direct computation of the overlap reduction function. As another example, it is easy to see that in GR, as is well-known~\cite{Taylor:2021yjx}, the dominant modes are the quadrupole {\it and} octopole modes, explaining why the minimum of the overlap reduction function takes place at $\xi_{\text{min}}^{\rm GR}\approx 82^\circ$, rather than at exactly $90^\circ$ as one would expect from only the quadrupole contribution. In fact, this behavior leads to one of the main results of this paper, namely that the position of the minimum of the overlap reduction function provides an interesting measure for departures from GR in PTA observations, as we shall see in the next section.

\section{Computing the Overlap Reduction Function Beyond GR: Examples}\label{sec:example}

In this section, we turn to examples in which the physics is different from that of GR, explicitly considering the overlap reduction function Eq.\eqref{eq,gammasum} in examples with $v_{ph}>1$ and $v_{ph}<1$ respectively.  
LIGO-Virgo observations~\cite{LIGOScientific:2017zic} constrain the group velocity of gravitational waves to satisfy $|v_g-1|\lesssim 10^{-15}$. In general, how this translates into a constraint on the phase velocity depends on the dispersion relation. 
Moreover, the constraint derived from measurements in the LIGO frequency band does not, in principle, forbid the deviation of the phase velocity from the speed of light in lower frequency bands, such as the nHz band relevant to PTA observations. It is therefore worthwhile considering signatures of deviations of the phase velocity from the speed of light in PTA observations.

We focus on the effects of these different dispersion relations on the tensor modes. In particular, we point out that the minimum angle of the overlap reduction function shifts due to the modification of the dispersion relation, and for certain cases is distinguishable from the shift due to the inclusion of extra polarization modes. This provides us with a new method with which to potentially identify deviations from GR from future PTA datasets.

\begin{figure}[h!]
\centering
\includegraphics[scale=0.6]{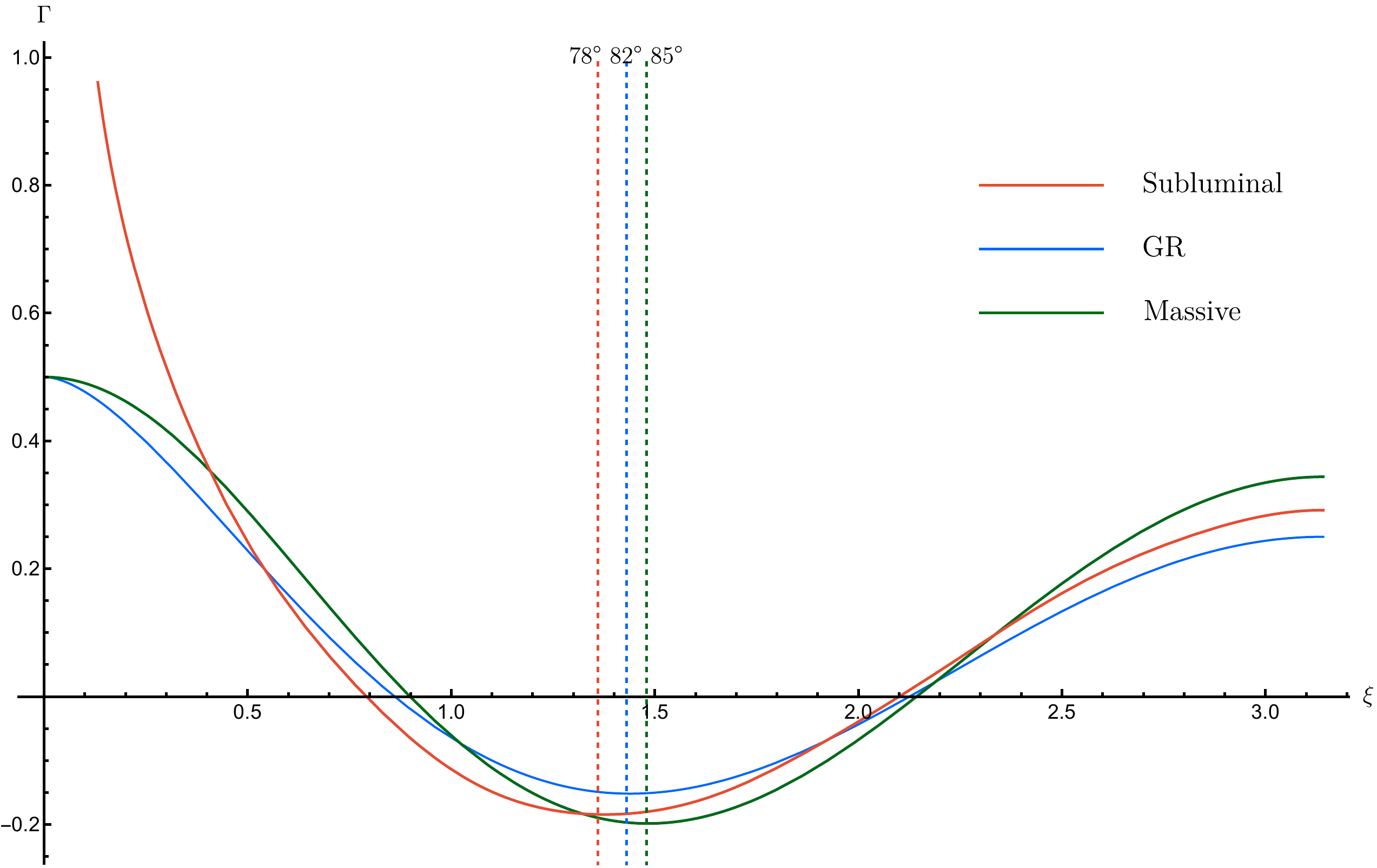} 
\caption{The overlap reduction function $\Gamma(\xi)$ in the $fL\gg 1$ limit for the cases of massive gravity with $1/v_{ph} = 0.9$ (green); gravity with a subluminal phase velocity $1/v_{ph} = 1.1$ (red); and GR (blue, the Hellings-Downs curve). 
The vertical dashed lines indicate the location of the minimum in each case.
One can see that the minimum in the massive gravity case has been shifted to the right; and the minimum for gravity with a subluminal phase velocity has moved to the left. 
For GR and massive gravity, we normalize the overlap reduction function such that at $\xi =0$, the value is chosen to be $0.5$. For the case with subluminal phase velocity, the overlap reduction function diverges at $\xi = 0$, as been discussed in Eq.\eqref{eq,divergence}. Therefore we choose an arbitrary normalization for comparison.
} 
\label{fig:massiveGamma}
\end{figure}  

\subsection{Massive Gravity} 
As been discussed previously~\cite{Liang:2021bct,deRham:2016nuf,Shao:2020exw,Bernardo:2023mxc}, the current bound on the mass of the graviton in massive gravity is around the scale to which PTA measurements are sensitive. It is therefore interesting to consider massive gravity as an example of the superluminal phase velocity case.

For massive gravity, we have the dispersion relation
\begin{equation}
    \omega(k) = \sqrt{k^2+m^2},
\end{equation}
where $m$ is the mass of the graviton. The phase velocity is then given by $v_{ph} \equiv \omega /k = \sqrt{k^2+m^2}/k>1$.

\begin{figure}[h!]
\centering
\includegraphics[scale=0.7]{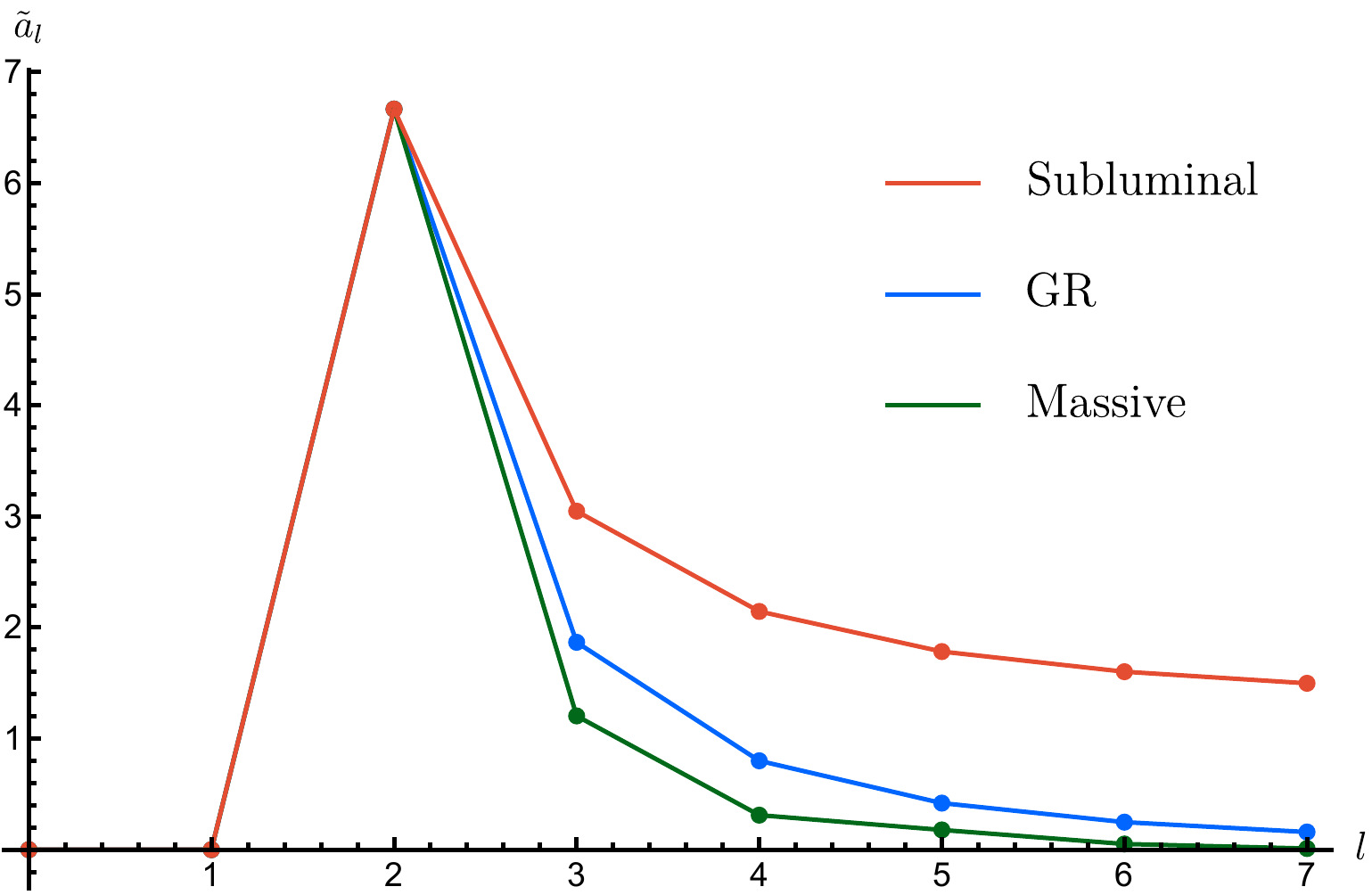} 
\caption{The normalized coefficients $\tilde a_\ell$ of the Legendre polynomial modes, in the $fL\gg 1$ limit, for GR (blue), massive gravity (green), and gravity with a subluminal phase velocity (red). For the benefit of making the enhancement/suppression clear, we normalize the coefficients $a_\ell$ in Eq.~(\ref{eq, overlapnew}) such that for quadrupole modes, the coefficients $\tilde a_\ell$ are identical. Note that the lines overlap for $\ell\leq 2$.}
\label{fig:coefficient}
\end{figure}

As an example, we show the overlap reduction function with $1/v_{ph}= 0.9$ in Fig.~\ref{fig:massiveGamma}. The Hellings-Downs curve in GR is also shown for comparison. We see that the location of the minimum angle shifts to the right in the massive gravity case compared to GR. This shift is a direct consequence of the suppression of the high multipole modes. 
In Table.\eqref{table:coeffcl1} and Fig.~\ref{fig:coefficient} we show numerical results for $|c_\ell|^2$ and $a_\ell$ for the first six multipole modes. One can see that in the case of massive gravity, the coefficients rapidly converge to zero, and the higher multipoles are further suppressed compared to GR. Therefore, the quadrupole mode becomes more dominant, and hence the minimum angle shifts towards $90^\circ$, i.e. to the right compared to the GR value $\xi_{\rm min}^{\rm GR}\approx 82^\circ$.
\vspace{5mm}

\begin{table}[ht]

\begin{tabularx}{0.8\textwidth} { 
    | >{\raggedright\arraybackslash}X 
    | >{\centering\arraybackslash}X 
    | >{\centering\arraybackslash}X 
    | >{\centering\arraybackslash}X 
    | >{\centering\arraybackslash}X 
    | >{\centering\arraybackslash}X
    | >{\centering\arraybackslash}X|}
\hline
& $\ell=2$ & $\ell=3$ & $\ell=4$ & $\ell=5$ & $\ell=6$ & $\ell=7$ \\
\hline
GR  & 4  & 4  & 4 &  4 & 4 & 4  \\
\hline
$1/v_{ph}$=0.9 & 3.74076   & 3.00424 & 2.33123 & 1.7623 & 1.30515 & 0.95075  \\
\hline 
$1/v_{ph}$=1.1   & 4.42402  & 5.65253 &7.24344  & 9.12632 &11.25257 &13.58964 \\
\hline 
\end{tabularx}
\caption{The coefficients $|c_\ell|^2$ of first six modes of Legendre polynomial expansion~(\ref{eq, overlapnew}), in the $fL\gg 1$ limit, for GR, massive gravity, and the case of a subluminal phase velocity. One can see that for higher $\ell$, the coefficients are constant for GR, suppressed in massive gravity case, and enhanced in the case with subluminal phase velocity. 
}
\label{table:coeffcl1} 
\end{table}

This rightward shift of the minimum angle is a generic feature for any $v_{ph}>1$ dispersion relation. To see this more clearly, we take the non-relativistic limit in massive gravity, in which $m^2 \gg k^2$, so that $v_{ph} \gg 1$. As was pointed out in \cite{Liang:2021bct}, for massive gravity it is a reasonable approximation to drop the exponential factor to obtain the analogous quantity to the Hellings-Downs curve. 
Setting the exponential factor in Eq.\eqref{eq,coeff} to unity, we can write
\begin{equation}\label{eq:clexpanded}
    c_\ell \approx  \int_{-1}^1 dx \frac{1-x^2} { 1+  x/v_{ph} } (1-x^2) \frac{d^2}{dx^2 } P_\ell(x)\ .
\end{equation}
In the $v_{ph}\gg 1$ limit, we can then expand the denominator as a power series in $1/v_{ph}$ to obtain
\begin{eqnarray}
\label{eq,clstation}
    c_\ell   &=& \sum_{n=0}^{\infty}     \int_{-1}^1 dx  \left(-\frac{x}{v_{ph}}\right)^{n}  (1-x^2)^2  \frac{d^2}{dx^2 } P_\ell(x ) \nonumber\\
    &=&   \int_{-1}^1 dx \left[ 8P_2(x) - \frac{8}{v_{ph}}P_3(x)  \right]P_l(x) \nonumber\\
    &+&\sum_{n=2}^{\infty}  \int_{-1}^1 dx   \frac{1}{v_{ph}^n}  \left(-x\right)^{n-2 } \left[ (n+3)(n+4)x^4 - 2(n+1)(n+2)x^2 +n(n-1) \right]   P_\ell(x ) \ ,
\end{eqnarray} 
where in the second and third lines we have integrated by parts and neglected boundary terms. 
Note that the terms in those lines take the form of Legendre Polynomials multiplied together, and so by the orthogonality relation the terms in the second line only contribute to the quadrupole and octopole modes, and the contribution to higher multipole modes comes from the terms in the third line. If we only consider the leading order contribution to each multipole mode, it is clear that the coefficients of all higher modes $|c_\ell|^2$ are suppressed by a factor of $1/v_{ph}^{2(\ell-2)}$ compared to GR. This is consistent with the results obtained by directly integrating Eq.\eqref{eq,gammaint}
\begin{eqnarray}
   \Gamma  = \frac{\beta_T }{4}\left[\frac{16\pi}{15}P_2(\cos\xi)  +\frac{16\pi}{105} \frac{1}{v_{ph}^2} ( 2P_2(\cos\xi)+P_3(\cos\xi)  )\right] + \mathcal{O}\left( \frac{1}{v_{ph}^4} \right) \ ,
\end{eqnarray} 
where we can see that the octopole mode is suppressed by a factor of $1/v_{ph}^2$.

The shift of the minimum angle depends on the deviation of the phase velocity from the speed of light or, equivalently, the graviton mass.
In the extreme case $v_{ph}\rightarrow\infty$ the overlap reduction function is completely dominated by the quadrupole mode and hence the minimum angle $\xi_{\rm min}^{v_{ph}\rightarrow\infty}=90^\circ$.

\subsection{Gravity with a Subluminal Phase Velocity }
\label{subsec:sublum}
We now turn to a different regime, in which the dispersion relation for the graviton is given by
\begin{equation}
\label{eq:DR-subvp}
    \omega = c_s k \ ,
\end{equation}
with the sound speed $c_s<1$, so that the phase velocity satisfies $v_{ph}=c_s<1$. Models exhibiting this behavior can arise, for example, when considering the effective field theory of gravity and including higher curvature terms (See, e.g. \cite{CarrilloGonzalez:2022fwg, Ezquiaga:2021ler,deRham:2019ctd}).  We shall see below that in this case the minimum of the overlap reduction function generically takes place at a lower value of the angle compared to its location in GR.

In the previous cases we have studied, and also for GR, it is a reasonable approximation to set the exponential factor in the overlap reduction function to unity. However, when generalized to the subluminal case one notices that, after dropping the exponential factor, there is an intrinsic singularity at $\hat \Omega\cdot \hat p = -v_{ph}$ in Eq.\eqref{eq,gammaint}, since $v_{ph}<1$ now. Fortunately, this singularity is not physical since, if we keep the exponential and expand about this point, then the pole in the denominator is canceled:
\begin{eqnarray}
    \lim_{\hat \Omega\cdot \hat p \approx  -v_{ph} } \left(\frac{ e^{-i 2 \pi f L\left(1+\frac{1}{v_{ph}} \hat{\Omega} \cdot \hat{p}\right)}-1 }{ 1+\frac{1}{v_{ph}} \hat{\Omega} \cdot \hat{p} } \right) \hat p^a \hat p^b e^{+,\times}_{ab}  = (-2\pi i f L ) \hat p^a \hat p^b e^{+,\times}_{ab} + \mathcal{O}\left(1+ \frac{1}{v_{ph}} \hat{\Omega} \cdot \hat{p} \right)\ .
\end{eqnarray}
Nevertheless, naively dropping the exponential leads to an obstacle to performing the kind of straightforward evaluation that was possible in the GR limit.

\begin{figure}[h!]
\centering
\includegraphics[scale=0.7]{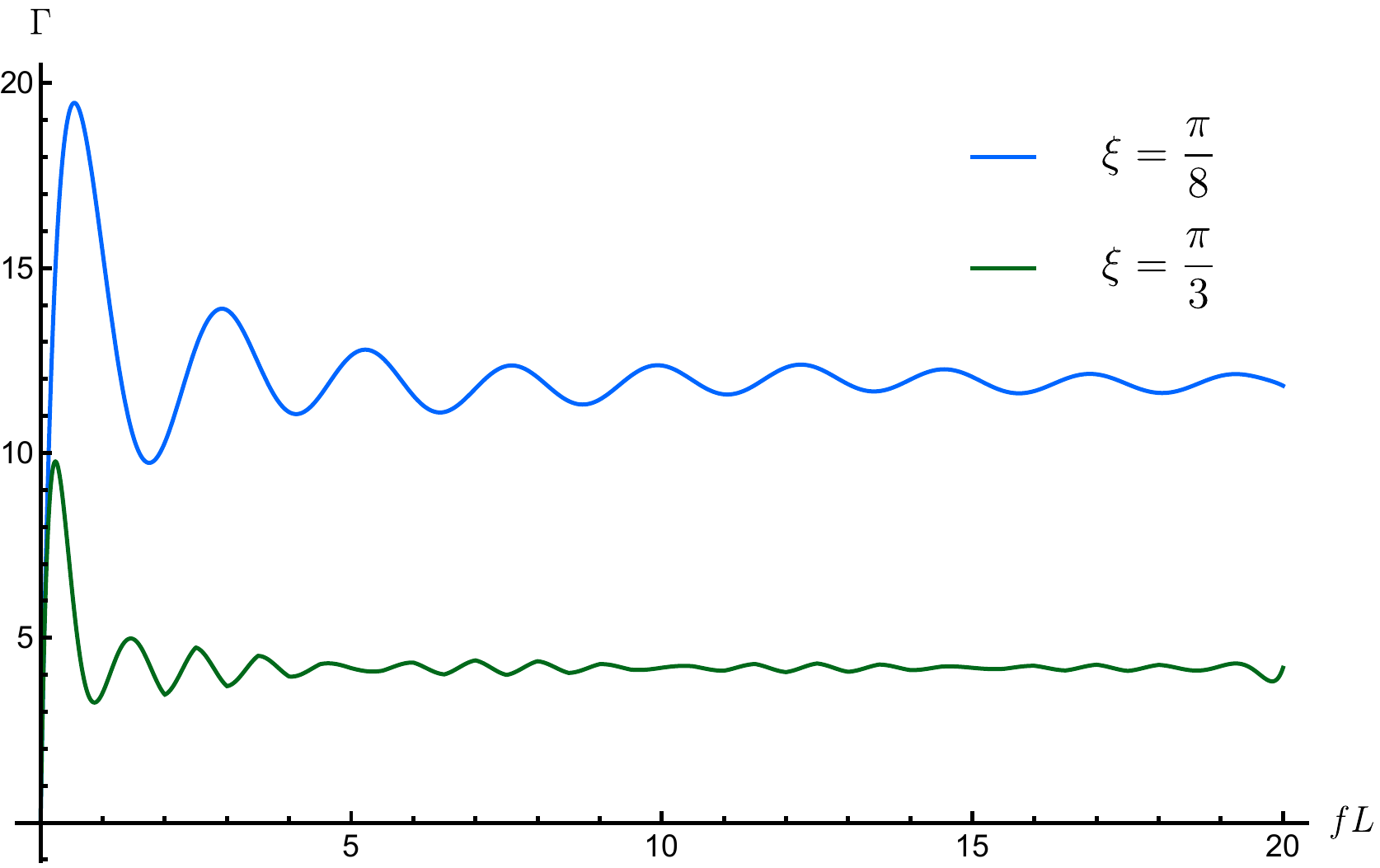} 
 \caption{Overlap reduction function obtained from direct numerical integration of Eq.\eqref{eq,gammaint} with $1/v_{ph}= 1.1 $ . Two choices of angle ($\xi = \pi/8$ (blue) and $\xi = \pi /3$ (green)) are chosen to show that the exponential factor plays the role of damping the oscillations.   
}
\label{fig:subluminalGamma}
\end{figure}

Although it remains clear that the exponential still plays the role of a damping term (See Fig.~\ref{fig:subluminalGamma}), because we can no longer neglect it the harmonic analysis is particularly useful for computing the angular distribution in this case. This technique allows us to simplify the question by expanding the overlap reduction function in a series of Legendre polynomials, and then numerically computing their coefficients. 

An important point here is the convergence behavior of the Legendre polynomial expansion at large $\ell$. We first rewrite $c_\ell$ in Eq.\eqref{eq,coeff} in the following form:
\begin{eqnarray}
    c_\ell = (\ell+1)2 i  \int_{-1}^1 dx  e^{-\pi i fL (1+ x/v)} \frac{\sin \pi fL (1+ x/v)}{(1+ x/v) } \left( (-\ell+(2+\ell)x^2)P_{\ell}(x) - 2x P_{\ell+1}(x)\right)\ ,
\end{eqnarray}
where we have used the Legendre differential equation, and have employed the recursion relation to replace $\frac{d^2}{dx^2} P_\ell(x) $. Note that, since we are interested in the large $fL$ limit, the familiar form of the $\delta$ function
\begin{eqnarray}
    \delta(x) = \lim_{\epsilon\to 0} \frac{\sin (x/\epsilon)}{\pi x}\ ,
\end{eqnarray}
 can be identified. Taking this limit, we have 
\begin{eqnarray}
    \lim_{fL \to \infty} c_\ell &=& (\ell+1)2\pi i  v \int_{-1}^1 dx  e^{-\pi i fL (1+ x/v)}   \delta(x+v)  \left( (-\ell+(2+\ell)x^2)P_{\ell}(x) - 2x P_{\ell+1}(x)\right)\ \nonumber\\
    &=& (\ell+1)2\pi i  v  \left( (-\ell+(2+\ell)v^2)P_{\ell}(-v) + 2v P_{\ell+1}(-v)\right) 
\end{eqnarray} 

\begin{figure}[h!]
\centering
\includegraphics[scale=0.5]{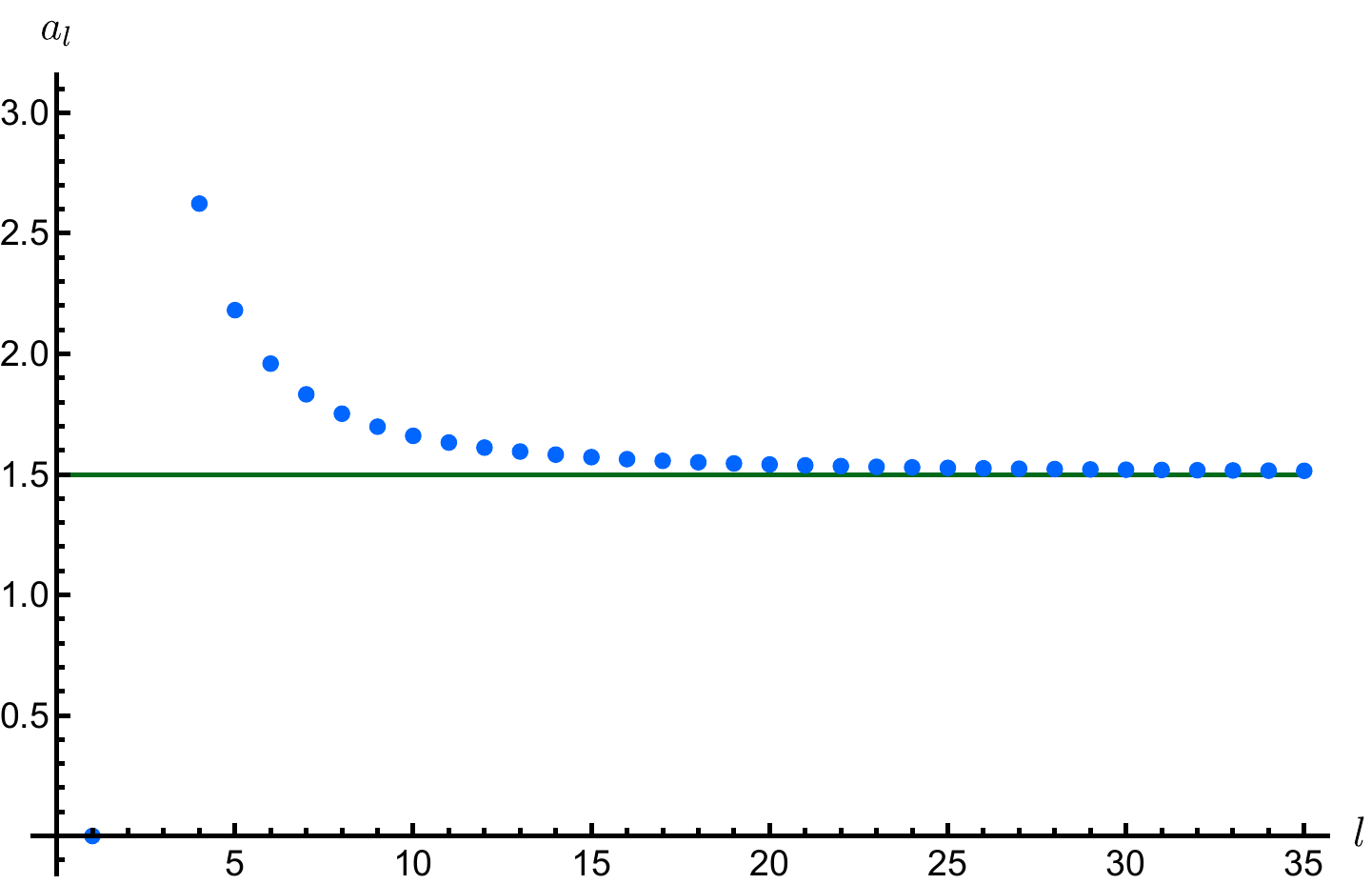} 
\includegraphics[scale=0.7]{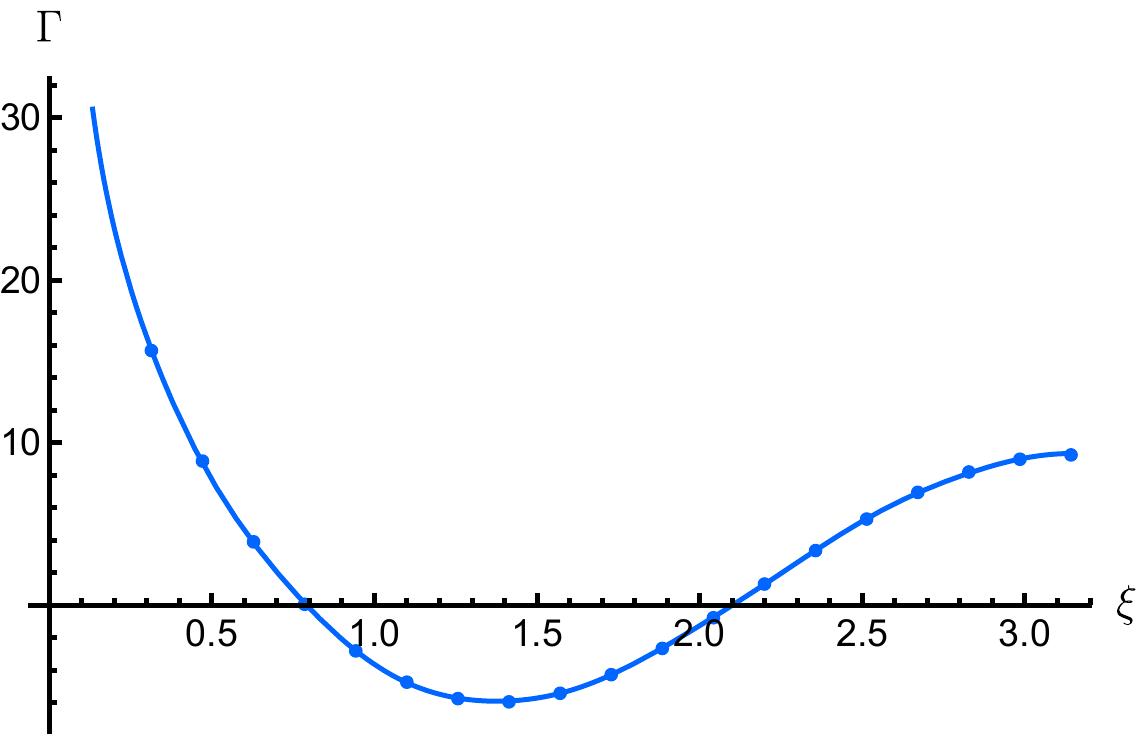} 
\caption{Left: The coefficients $a_\ell$ in the Legendre polynomial expansion for gravity with subluminal phase velocity $1/v_{ph}=1.1$. One can see that $a_\ell$ is approaching a constant, 1.5 in this case, at large $\ell$. Right: Overlap reduction function of gravity with the same subluminal phase velocity model computed as a summation of Legendre polynomials from $\ell=2 $ to $\ell=29$. The dots are the points obtained directly from the integration of Eq.\eqref{eq,gammaint} for different angles. 
}
\label{fig:coefflistplot}
\end{figure}

We then make use of the asymptotic behavior of the Legendre polynomials for large $\ell$ 
\begin{eqnarray}
    P_\ell(\cos\theta) =  \frac{2 }{\sqrt{2\pi \ell \sin\theta}}\cos\left(\ell+\frac{1}{2}\right)  \theta +\mathcal{O}\left(\ell^{-3 / 2}\right) 
\end{eqnarray}
to see that the coefficient $c_\ell$ behaves like $\ell^{3/2}$ at large $\ell$. The coefficient of the Legendre polynomial $a_\ell $ defined in Eq.\eqref{eq, overlapnew} therefore approaches a constant at large $\ell$. Now, a well-known Legendre identity is:
\begin{eqnarray}
\label{eq,divergence}
    \frac{1}{\sqrt{2}\sqrt{1-\cos\xi}} = \sum_{\ell = 0}^\infty P_\ell(\cos\xi)\ .
\end{eqnarray}
Subtracting this from the overlap reduction function, and adding terms back in to compensate for the difference at small $\ell$, we obtain a useful truncation, since $a_\ell$ approaches a constant rather rapidly. In the left-hand panel of Fig.~\ref{fig:coefflistplot} we numerically evaluate the $a_\ell$ and see that our approximation is valid, since they indeed approach a constant ($1.5$ in this case, with $1/v_{ph}=1.1$) when $\ell \gtrsim 10$. In the right-hand panel we then further verify our multipole truncation by comparing it to the direct numerical integral. 
Notice the divergence when $\xi\approx 0$, which can also be seen from Eq.\eqref{eq,divergence}. This occurs because the non-vanishing constant coefficients of the Legendre polynomial imply that we have contributions from infinitely many multipole modes at $P_\ell (1) = 1$.  Similar divergences also happen in other modified gravity theories with longitudinal scalar modes (See, e.g.~\cite{Qin:2020hfy,PhysRevD.85.082001}). 

Compared to GR, we find that the minimum correlation angle for a subluminal phase velocity is shifted towards the left, $\xi_{\text{min}} = 78.87^\circ$, as shown in Fig.~\ref{fig:massiveGamma}. This comes from the enhancement of the higher multipole modes as shown in Fig.~\ref{fig:coefficient}.  Notice that this is a distinct signature that cannot be explained by extra polarization modes, since these only move the minimum angle toward the right~\cite{Qin:2020hfy,PhysRevD.85.082001,Gair:2015hra}.

\section{Conclusions and Discussions}\label{sec:conclude}
In this paper we have considered the capacity for testing gravity using future detections of the stochastic gravitational wave background by pulsar timing array experiments. In particular, we have shown that angular correlations in the SGWB can serve as an effective diagnostic for deviations from GR. Assuming that the SGWB can be approximated as plane waves, we have shown that the modification to the angular correlation depends on the phase velocity of GWs. This is particularly interesting, since observations of the gravitational wave signal from binary coalescences constrain the {\it group} velocity of GWs, so that the SGWB in principle provides an alternative probe of gravity. 

We have studied this in detail for the case of massive gravity, for which the phase velocity is superluminal, and for the case in which there is a subluminal phase velocity. Compared to GR, a lower (higher) phase velocity increases (decreases) the contribution of the higher multipoles, hence changing the shape of the overlap reduction function. Specifically, the minimum angle of the overlap reduction function shifts to a larger (smaller) value due to a lower (higher) phase velocity.  A larger shift of the minimal correlation angle is particularly interesting, since contributions from scalar and vector modes always lead to a smaller shift. These features can be used to distinguish among theories of gravity in future PTA observations.

Taking into account the noise in the measurements, the cosmic variance of the overlap reduction function is smaller at the minimum angle $\xi_{\rm{min}}$ than at its value as $\xi \rightarrow 0$ or $\xi \rightarrow \pi$ \cite{Allen:2022dzg,Bernardo:2022xzl} (We thank the authors of \cite{Bernardo:2022xzl} for pointing this out to us.). This further emphasizes the advantages of using the shift of the minimum angle to discriminate among models. Furthermore, since the cosmic variance reaches its minimum around the zeros of the overlap reduction function, another natural possibility is to consider the shift of these zeros as another distinctive feature. If we only detect tensor modes with PTA measurements, then the shift of zeros is in the same direction as that of the minimum angle in the case of modified gravity. Thus, this shift of zeros could also be used to distinguish among different gravitational theories. However, if there are contributions from other polarization modes, for example a monopole contribution, then the direction of the shift of the zeros is more difficult to interpret.

\acknowledgements
We thank Wayne Hu and Austin Joyce for useful discussions. We are also grateful to Reginald Christian Bernardo and Kin-Wang Ng for important comments on the first draft of this paper. The work of QL and MT is supported in part by US Department of Energy (HEP) Award DE-SC0013528. M-X. L. is supported by funds provided by the Center for Particle Cosmology.

\bibliography{ref} 
\end{document}